\documentstyle[12pt]{article}
\begin{document}
\begin{titlepage}
\baselineskip 20pt

\title{Reaction $\gamma \pi \to \pi \pi$ in a Confined Quark Model}

\vspace{0.5cm}

\author{M. A. Ivanov\thanks{Permanent address: Bogoliubov Laboratory of
Theoretical Physics, Joint Institute for Nuclear Research, 141980 Dubna
(Moscow Region), Russia}
\, and T. Mizutani\\
\\
\\
Department of Physics\\
Virginia Polytechnic Institute and State University\\
Blacksburg, VA 24061 USA}
\maketitle

\abstract
\baselineskip 20pt
A confined quark model study in a couple of {\it chirally anomalous}
processes is presented
in comparison with effective meson Lagrangian approaches of various
kind. The processes considered are  $\pi^0 \to \gamma \gamma$
and $\gamma \to \pi^+ \pi^- \pi^0$ (or the equivalent $\pi \gamma \to \pi
\pi$) for which there is a well-known low energy theorem relating the
latter amplitude with the former one by a very simple algebraic relation in the
zero energy  (or chiral, or soft pion) limit.
Our quark model naturally generates the so-called
contact term in the amplitude for the second process, but with the opposite
sign to what effective chiral meson models indicate. A  reinterpretation
of our vector pole contribution restores the consistency, however.
While the first reaction is
observed to serve in {\it calibrating} various models, it is found difficult,
based upon the quality
of the existing data in the second reaction, to single out the {\it best
model}, which appears indispensable for testing the validity of the above
low energy theorem. Thus
the proposed experiments and their analyses should aim at attaining an
optimal accuracy.

\bigskip
\noindent PACS number(s): 11.30.Rd, 11.40.Ha, 12.40.Vv, 13.60.Le.
\end{titlepage}

\section{Introduction}
\baselineskip 20pt

A little more than a quarter of a century ago, the chiral (axial) anomaly was
found to amend the ordinary PCAC (partial conservation of axial current)
relation in order to account for the physical
decay of the neutral pion into two gammas, which otherwise would be too small
and vanish in the strict chiral (zero meson mass) limit \cite{Adletet}. As
has been frequently quoted, a quantitative comparison of the theoretical
(anomalous PCAC) prediction with the measured decay rate lead to one of the
strongest evidence for the three-colour nature endowed to the quarks. More
precisely,
together with a proportionality constant $N_c$ the anomalous PCAC gives the
$\pi^0 \to \gamma \gamma$ amplitude

\begin{equation}
F_{\pi\gamma\gamma}=\alpha N_c/3 \pi F_\pi = 0.025
\; {\rm GeV^{-1}},
\end{equation}
where $F_\pi  \approx 93$ MeV is the charged pion decay constant. In
comparison with the corresponding quantity  extracted from experiment:
$F_{\pi \gamma \gamma}^{\rm exp} = (0.0250 \pm 0.0003)$ GeV$^{-1}$, one obtains
$N_c = 3$: the number of colours, see for example \cite{Bijn0}.

A couple of years later, a low energy theorem which relates the
$F_{\pi \gamma \gamma}$ to the amplitude for the
$\gamma \to 3 \pi$ reaction at zero energy (more precisely the soft pion and
soft photon limit) which we shall denote as $F_{3\pi}(0) \equiv
F_{3\pi}(0,0,0)$ was deduced based
upon the
anomalous PCAC and current algebra by Adler {\it et al.} \cite{Adler},
Terent'ev
\cite{Teren}, and Aviv and Zee \cite{Aviv}: henceforth we shall call this
low energy theorem as the ATA theorem for brevity.  The occurrence of this
$\gamma \to 3\pi$ was also recognised as due to chiral anomaly, and the
theorem claims

\begin{equation}
eF_{3\pi}(0) = F_{\pi \gamma \gamma}/F_\pi^2,
\end{equation}
or
\begin{equation}
F_{3\pi}(0) = \frac{e}{4 \pi^2 F_\pi^3} \equiv F_{3\pi}^{\rm anom} =
9.54 \; {\rm GeV^{-3}}.
\end{equation}
Note that the value of the chiral anomaly prediction denoted as
$F_{3\pi}^{\rm anom}$ in the above equation will appear frequently in what
follows. Later it was
shown that, from the electromagnetically gauged
Wess-Zumino-Witten (WZW) term \cite{Wess}\cite{Witten} involving the
chiral fields (ordinary pseudo-scalar octet mesons), both
 $F_{\pi \gamma \gamma}$ and $F_{3\pi}^{\rm anom}$ can be obtained as the
tree-level amplitudes explicitly
satisfying the ATA theorem almost
for free \cite{Witten}\cite{Kay}\cite{Pak}.

To test the ATA theorem,
several physical processes like $\pi \to 2\pi$ in the Coulomb field
of heavy nuclei (Primakoff effect), $e^+ e^- \to e^+ e^- 3\pi$, etc. were
suggested for experiments \cite{Teren} \cite{Zee}.  In order then to
either deduce the zero energy (the soft pion limit) amplitude from
experiments or to predict
appropriate cross sections from the theoretical side,
it should be necessary to know how the appropriate physical amplitude may
be extrapolated to the zero energy limit, or conversely, how the zero energy
amplitude be continued to the physical region. The original
(electromagnetically gauged) Wess-Zumino-Witten term alone does not really
serve this objective and the standard procedure was to assume the process
to go via the formation of vector mesons: $\rho$ (and $\omega$)
\cite{Teren}\cite{Rudaz}. Note that this procedure was not
unique as will be discussed.
Two experimental results have been available: one from the Primakoff effect
with 40 GeV pion beam \cite{Antip}, see Fig.1, and the other from the
$\pi^- e \to
\pi^- \pi^0 e$ with 300 GeV pions at the CERN SPS \cite{Amend}, Fig.2.
For both
cases, all different methods explored in those articles for extrapolation
from zero energy amplitude fell inside the experimental error bars (about
 20$\%$ of the
central values) once the number of colours $N_c$ was fixed at three.
So one might feel that at least the colour problem has found the
solution, but that the ATA theorem has
not yet been proven:
both the theoretical models for describing the $\gamma \to 3\pi$  in
the zero energy limit and at finite energies and the experiments should be
improved.
In fact, there have been various effective meson Lagrangian models
constructed so as to be consistent with the underlying QCD
(including the now fashionable Chiral Perturbation Theory [CHPT], see for
example \cite{Gasser}). And on the
experimental side (i) there are plans to repeat the pion Primakoff process
with 600 GeV pions at Fermi Lab. (Fig.1), see for example \cite{Moin},
expecting a better
statistics and making the virtual photons
to become closer to the real ones than those in the previous
experiment  \cite{Antip}, and (ii)
an accepted proposal (at CEBAF) for  studying $\gamma \pi^+ \to
\pi^+ \pi^0$ in the  photo double pion production: $\gamma p \to \pi^+
\pi^0 n$ (Fig.3), see \cite{CEBAF} using the tagged photon beam with energies
between 1 and 2 GeV  (note that this experiment was investigated before
 \cite{Mesh}, but the result was strongly in contradiction with the
consequences of both
refs. \cite{Antip} and \cite{Amend}). So both theoretical and
experimental
activities are expected to shed a new light on testing the ATA theorem
with improved precisions.

In what follows we shall first discuss some salient features of various
effective meson Lagrangian approaches for the $\gamma \pi \to 2\pi$
(or $\gamma \to 3\pi$) reaction and related processes in Section 2,
 particularly the ways in which the vector meson contributions are
implemented.  This section may appear a little lengthy, but we consider it
useful to give some overview of the theoretical trend in a comprehensible
manner. Incidentally, by this one might see that
Chiral Perturbation Theory (CHPT) is $not$ the only relevant approach to low
energy meson phyiscs. Then in
Section 3, we present our confined quark approach for the same reaction.
This we think should be considered as an alternative to those effective
Lagrangian approaches discussed in Section 2 where only meson degrees of
freedom are the relevant quantities
(with some minor exception of the appearance of current quark mass ratios,
quark electric charges, etc.).
Section 4 is devoted to an explicit calculation of cross section
$\sigma/Z^2$ for the Primakoff process: the pion production in a nuclear
Coulomb field on by an incident pion,
from some models including our own, and compare with the experimental result
in \cite{Antip}. Then a short conclusion will be given at the end.

\section{Models for Vector Meson Contributions: a Brief History}
\baselineskip 20pt

As mentioned in the previous section, chiral anomaly predicts the zero energy
$\gamma \to 3\pi$ amplitude  $F_{3\pi}(0)$ which is a real constant.
It must be continued
to appropriate {\it physical regions} in order to be confronted with
relevant quantities
extracted from experiments (or the continued amplitude must be used to
calculate experimentally measurable quantities). The continuation
(or extrapolation)
then makes the amplitude a function of several Lorentz invariant
kinematical variables (the number reduces to three, of course, when all the
particles are physical, of which only two are independent). It was assumed
that the continuation from zero energy would arise due to the
$\rho$ and $\omega$ vector meson pole contributions for reasonable
values of these kinematical invariants, say up to
$\sim 1 \; {\rm GeV^2}$ \cite{Teren}. For the discussion to follow,
we choose the process $\gamma \pi \to \pi \pi$  to define the kinematical
variables, see Figs. 1-2.
In the physical region for this process, $s>4m_\pi^2$ and $t<0$ should hold.
Note that in the experiments in \cite{Antip} \cite{Amend}, the photon is
reasonably close to the real one: $-q^2<2\cdot 10^{-3}$ GeV$^2$ \cite{Antip},
and
$-q^2\approx 10^{-3}$ GeV$^2$ \cite{Amend}. For the proposed experiment
\cite{CEBAF}, the photon is real: $q^2 = 0$ but the incident pion with
momentum $p_1$ is virtual: $p_1^2 < 0$.
Mandelstam variables are defined as

$$
  s=(p_1+q)^2=(p_2+p_3)^2, \hspace{0,5cm}
  t=(p_1-p_2)^2=(p_3-q)^2, \hspace{0,5cm}
  u=(p_1-p_3)^2=(p_2-q)^2,
$$
so that for processes of \cite{Antip} and \cite{Amend} \  $s+t+u=3m_\pi^2+q^2.$
With this preparation we shall give a review of several existing models
predicting the zero energy (all the particles are {\it soft}: $s=t=u=q^2=0$)
 amplitude and its continuation
to physical regions.
As stated in the Introduction, ATA \cite{Adler}\cite{Teren}\cite{Aviv} found
 $
F_{3\pi}(0)=F_{3\pi}^{\rm anom}$
from anomalous PCAC and current algebra. Then Terent'ev \cite{Teren}
devised a simplest continuation from the chiral anomaly prediction in the
form
\begin{equation}
F_{3\pi}(s,t,u)= F_{3\pi}^{\rm anom} \biggl[1+C_\rho e^{i\delta}
\biggl(\frac{s}{m_\rho^2-s}+\frac{t}{m_\rho^2-t}+\frac{u}{m_\rho^2-u}\biggr)
+C_\omega e^{i\delta'}\frac{q^2}{m_\omega^2-q^2}\biggr],
\end{equation}

$$
C_\rho=\frac{2g_{\rho\pi\pi}g_{\rho\pi\gamma}}
            {m_\rho^3F_{3\pi}^{\rm anom}}\approx 0.434, \hspace{1cm}
C_\omega=\frac{eg_{\omega\gamma}g_{\omega 3\pi}}{m_\rho^3F_{3\pi}^{\rm anom}}
\approx
3,
$$
where $\delta$ and $\delta'$ are unknown relative phases.  This is a form
inferred purely intuitively from the assumptions of (i) the vector meson
dominance (hereafter denoted as VMD), and (ii) $F_{3\pi}(s,t,u)$ satisfying
a once-subtracted dispersion relation with  $F_{3\pi}^{\rm anom}$ the
subtraction
constant at zero energy.

The second approach to be mentioned here is the work of Rudaz \cite{Rudaz}
based upon the pole model of Gell-Mann {\it et al.} \cite{Gell} in which
VMD is the guiding principle: all the radiative meson processes go via
productions of vector mesons. First, Rudaz obtained the
$F_{\pi \gamma \gamma}$ using the SU(3), $\rho$
universality and the $\omega \to \pi^0 \gamma$ width and found an agreement
with the prediction from chiral anomaly.  Then he calculated the
$\gamma \to 3\pi$ amplitude and found that without anomalous PCAC the
ATA theorem could be obtained only
if the revised Kawarabayashi-Suzuki-Riazuddin-Fiazuddin (KSRF) relation
\cite{KSRF} would be adopted (\` a la Basdevant and Zinn-Justin (BZ)
\cite{BZ}) to relate the vector meson and pion quantities at low energies: by
 defining
\begin{equation}
\alpha_k \equiv \frac{g^2_{\rho \pi \pi}F_\pi^2}{m_\rho^2},
\end{equation}
with $m_\rho$ and $g_{\rho \pi \pi}$  the $\rho$ meson mass and universal
coupling constant (the universality means, for example, $g_{\rho \pi \pi} =
g_{\rho NN}$)
respectively, the KSRF relation fixes its value as  $\alpha_k  = 1/2$
 (its current empirical value is about $0.53$)  whereas
BZ obtained $\alpha_k = 1/3$. With this latter value the
resultant Rudaz amplitude reads
\begin{equation}
F_{3\pi}(s,t,u)= F_{3\pi}^{\rm anom}
\frac{m_\omega^2}{m_\omega^2-q^2}
\frac{m_\rho^2}{3}
\biggl(\frac{1}{m_\rho^2-s}+\frac{1}{m_\rho^2-t}+\frac{1}{m_\rho^2-u}\biggr).
\end{equation}
As a function of $s$ the ratio
$F_{3\pi}(s,0,0)/F_{3\pi}^{\rm anom}$ from the Terent'ev
(with unknown phases set equal to zero) and the Rudaz models are not
identical but the difference is seen as
at most $10\%$, which might roughly define what kind of
experimental accuracy would be required if one wished to distinguish the model
difference from given data.

The story became quite interesting when Rudaz later claimed \cite{Rud2} that
in view of the later development the above BZ choice:
$\alpha_k =1/3$, would be
somewhat dubious, and that the KSRF choice: $\alpha_k =1/2$ must be adopted.
Then in order to respect the ATA theorem
                  he suggested to introduce  a  non-vanishing direct contact
term for the
$\omega\to 3\pi$ coupling with its value to be fixed at
$G_\omega =-g_{\rho \pi \pi}/16 \pi^2 F_\pi^3$,
in the zeroth order in $m_\pi^2/m_\rho^2$. This led to the
 revised Rudaz amplitude:
\begin{equation}
F^c_{3\pi}(s,t,u)= F_{3\pi}^{\rm anom}
\frac{m_\omega^2}{m_\omega^2-q^2}
\biggl[-\frac{1}{2}+\frac{3}{2}\frac{m_\rho^2}{3}
\biggl(\frac{1}{m_\rho^2-s}+\frac{1}{m_\rho^2-t}+\frac{1}{m_\rho^2-u}\biggr)
\biggr].
\end{equation}
Thus the na\" {\i}ve pole contribution in VMD model alone appeared insufficient
and
an empirical rather large and negative direct $\gamma 3\pi$ contact term
contribution seemed indispensable. This result was later given a support
by Cohen \cite{Cohen} who recalled the structure of the anomalous
Ward identity for $\gamma \to 3\pi$ discussed by Aviv and Zee \cite{Aviv}
from which the sign and magnitude of the contact and vector meson pole
contributions for the above expression could be read off.

In relation to the forgoing discussion, it may be important here to touch
upon the subject of the possible non-zero
$\omega 3\pi$ contact term the history of which is rather old, but  still
lives in modern effective  meson Lagrangian approaches.
Earlier, in a pole dominance approximation to the
isospin one $\rho \pi$ scattering combined with the assumption of unsubtracted
dispersion relation \cite{Brown}, its possible existence was suggested once
a consistency with Bjorken limit was imposed. Later, there was an attempt to
determine its magnitude from the point of view of hard current algebra plus
anomalous Ward identity for three and four point functions in describing
the $\omega \to \pi^+ \pi^- \pi^0$ and $\gamma \pi \to 2\pi$ \cite{Ali},
but no definite value could be reached. Within the experimental accuracy
available \cite{Codier}, the $\omega \to 3\pi$ transition could be explained
by the pole model of VMD \cite{Gell} with no apparent need for the finite
contact term. So for a while the subject stayed dormant.

{}From mid-80's there arose interest in
constructing effective meson Lagrangian by adding vector (pseudo-vector) meson
fields to the combination of the ordinary pseudoscalar (chiral)
Lagrangian and the Wess-Zumino-Witten (WZW) term.
This was a revival of the once successful vector meson dominance model (VMD) in
a more refined form inferred from underlying QCD.
Of those we single out the works by Kaymakcalan {\it et al.} \cite{Kay},
Brihaye
{\it et al.} \cite{Pak}, and Kuraev  and Silagadze \cite{Kura}. Kaymakcalan
{\it et al.}
\cite{Kay}
tried a  non-abelian (vector and axial vector fields) gauging of the
chiral plus WZW action
and, after introducing the VMD (to couple the photon with vector mesons),
they required to arrive at what is called the
Bardeen form of the
anomaly. This was found to correctly predict the $\pi^0 \to 2\gamma$
decay rate through the $\omega \rho \pi$ coupling: a kind of prerequisite for
any sensible model.  The model has given a definite value for the
$\omega 3\pi$ contact term as a function of $\alpha_k$ as introduced before.
For the KSRF choice: $\alpha_k =1/2$, which has given about 1/4 of the Rudaz
value for the contact term, this model did successfully reproduce
the $\omega \to 3\pi$ rate but did not  furnish the ATA theorem
(in fact, Rudaz has stated \cite{Rud2} that for any
sensible value of $\alpha_k$ the low energy theorem is not reproduciable in
this model). Note, however, that as an apparent converse, the Rudaz value for
the contact term cannot supply the sufficient $\omega \to 3\pi$
rate consistent with experiment \cite{Kura}. It should be interesting to
remark in passing that the authors of ref. \cite{Kura} have demonstrated
that once the kaon loop correction with the WZW term for the
$K^+ K^- \to 3\pi$ contact interaction to  the $\gamma \to 3\pi$ amplitude is
added within the model of ref. \cite{Kay}, the ATA theorem can be restored
upon suitable regularisations, the consistency condition for which
makes $\alpha_k = 0.55$, very close to the empirical value
(the pion loop contribution is shown to be small $O(m_\pi^2)$, thus may be
disregarded).

The second set of models \cite{Pak} \cite{Kura} demanded only the U(1)
electromagnetic gauging of the combined chiral plus WZW
effective action, then tried to
consistently introduce vector mesons in an empirically well-established
principle of VMD. The models easily passed the test to reproduce the
$\pi^0 \to 2\gamma$ rate. Here again the parameter $\alpha_k$ enters and with
its value set equal to the KSRF one, the $\omega 3\pi$ contact term agrees
with the Rudaz one, so the ATA theorem was shown to be
satisfied (this means, however, that the prediction for the
$\omega \to  3\pi$ turned smaller than the empirical value).  Now summing up,
those and other effective Lagrangian approaches may need to be confronted
with other radiative meson processes, etc. and further relevant experiments
are desirable. We note in this regard that a discussion in \cite{Kramer}
should be quite useful.

Last in line to be discussed is the chiral perturbation theory (CHPT)
applied to $\pi \gamma \to 2\pi$ and $\eta \to 2\pi \gamma$ by Bijnens
{\it et al.}
\cite{Bij}. They started with the same chiral plus WZW action like in the
works discussed above \cite{Kay} \cite{Pak} \cite{Kura} and, calculated the
corrections  to $F_{3\pi}^{\rm anom}$ which, as has
already been stated previously, is obtained upon simply gauging the WZW term
by the photon field.
The corrections include one loop diagrams involving one vertex from
the WZW term, and tree diagrams from the $O(p^6)$ effective
Lagrangian appropriate for the process which is required also to cancel the
divergence from the
loop-WZW combination. The finite part of the $0(p^6)$
tree coefficients were fixed by assuming their saturation by the nonet vector
meson contributions and VMD, then by integrating over the vector meson fields
(so, needless to say, no explicit vector mesons but only the original
pseudoscalar octets appear in the effective Lagrangian). The resultant
correction term is shown to have the following form,

\begin{equation}
F^{CHPT}_{3\pi}(s,t,u)= F_{3\pi}^{\rm anom}
\biggl(1+C^{\pi^0}_{\rm loops}(s,t,u)+\frac{3m_\pi^2}{2m_\rho^2}\biggr),
\end{equation}
where $C^{\pi^0}_{\rm loops}(s,t,u)$ is the loop correction:

\begin{eqnarray}
C^{\pi^0}_{\rm loops}(s,t,u)&=&\frac{m_\pi^2}{32\pi^2 F_\pi^2}
\biggl\{
\ln{ \frac {m_\rho^2} {m_\pi^2} }-7
  - \frac {\bar s-4}{3} \sqrt{ \frac {\bar s-4} {\bar s} }
\ln{ \frac {1+\sqrt{1-4/\bar s} } {1-\sqrt{1-4/\bar s} } }
\biggr.
\nonumber\\
&&\nonumber\\
&& + \frac {\bar t-4}{3}  \sqrt{\frac{\bar t-4}{-\bar t} }
\ln{ \frac {\sqrt{1-4/\bar t}+1} {\sqrt{1-4/\bar t}-1} }
+\frac {\bar u-4}{3} \sqrt{\frac{\bar u-4}{-\bar u} }
\ln{ \frac {\sqrt{1-4/\bar u}+1} {\sqrt{1-4/\bar u}-1} }\nonumber\\
&&\nonumber\\
\biggl.
&&-i\pi \frac {\bar s-4}{3} \sqrt{\frac {\bar s-4} {\bar s} }
\biggr\}.
\nonumber
\end{eqnarray}

Here $\bar s=s/m_\pi^2$, $\bar t=t/m_\pi^2$, and $\bar u=u/m_\pi^2$.
The appearance of the $m_\rho$ in logarithm is a standard feature
arising from the mass scale needed for dimensional regularisation adopted:
a common practice in CHPT is to set this scale equal to the $\rho$ mass.
Needless to say that the $0(p^6)$ correction should vanish in the chiral
limit.  It may be of use to mention that the origin of the third term in the
big bracket in Eq.(8) due to the $O(p^6)$ tree contribution
is basically the second term in the linearised
vector meson propagator contributions, viz.
$$
\frac{m_\rho^2}{m_\rho^2 - s} \approx 1  +\frac{s}{m_\rho^2} +
O[(s/m_\rho^2)^2],
$$
together with $s + t + u = 3m_\pi^2$, ($q^2= 0$ for the physical photon).
This approximation may work typically up to
$|s, t \; {\rm or} \; u| \approx 8 m_\pi^2$: roughly the validity limit
of the CHPT expansion. For a physical  $\gamma \pi \to 2\pi$ process (where
$s \ge 4m_{\pi^2})$ this linearised version is therefore expected to behave
differently from the VMD results  discussed above beyond this limit for $s$,
even without
considering the loop contribution.

\section{A Confined Quark Approach to the $\gamma \to 3\pi$ Amplitude}
\baselineskip 20 pt

In the previous section we have seen  effective meson Lagrangian approaches
which were used not only to find the relation between
$F_{\pi \gamma \gamma}$ and $F_{3\pi}(0)$: to establish the ATA low energy
theorem,
but to continue the latter quantity into
appropriate physical regions.  All modern versions of these approaches
are based upon the sum of the lowest order chiral Lagrangian and the
corresponding Wess-Zumino-Witten (WZW) term in order to ensure the
possibility of
retaining chiral anomaly which allows transitions: even number of
pseudoscalar  mesons $\leftrightarrow$ odd number of pseudoscalar mesons.
 The way in which the photon and
vector (and axial vector) meson fields are introduced  to this combination
differs
in different models: some are rather close to each other while others are
not. But the basic
guideline (often employed in implementing the vector-axial vector fields)
is that once these fields are formally
integrated out, the resulting action functional is just the original
chiral plus WZW term gauged electromagnetically.  Thus in a way those
effective meson Lagrangians are  consistent with chiral anomaly
{\it by construction} and in one way or another (recall the kaon loop
correction \cite{Kura} for the model of ref. \cite{Kay}) should reproduce
the ATA theorem. In this regard we have been
curious about if there may be a way to approach our chirally anomalous
processes from the quark model point of view. Our interest in this regard is
not in those models like in \cite{Vol} or \cite{Tucci} in which
either specific quark models or QCD Lagrangian is adopted in attempting to
derive the effective (non-renormalisable) meson Lagrangian.  Rather, we want
to regard mesons as explicit quark-antiquark bound states, so photons
couple to mesons through quarks, and various electromagnetic and mesonic
transitions proceed explicitly
through quark loops. Thus our vector mesons are $not$ of Yang-Mills
(non-abelian gauge) type. Besides, no VMD assumption is present.
This may well give different form of amplitudes for the processes of
our interest here as compared with those from the
effective meson Lagrangians discussed so far. In principle, any quark model
which may be able to describe low energy meson properties will do, and here
we shall adopt a confined quark model: Quark Confinement Model developed at
Dubna (QCM). There is a detailed monograph \cite{EI} on this approach so
we only supply here the minimum prerequisite for the present purpose.

In QCM \cite{EI},
generally there are contributions both from quark loops and meson pole
diagrams. The
gauge invariance of the Lagrangian under electromagnetic interaction leads to
the
gauge invariant form of the transtions $V\to\gamma$ ($V$: vector meson)
that gives a vanishing contribution with real photons.
 The model is specified by the interaction Lagrangian
describing the meson transition into quarks, for example, for $\pi$ and
$\rho$ mesons one can write

\begin{equation}
L_I(x)=
\frac{ig_\pi}{\sqrt{2}}\vec\pi(x)\bar q(x)\vec\tau\gamma^5 q(x)+
\frac{g_\rho}{\sqrt{2}}\vec\rho^\mu(x)\bar q(x)\vec\tau\gamma^\mu q(x)+
eA^\mu(x)\bar q(x)Q\gamma^\mu q(x)
\end{equation}
with the meson-quark-quark coupling constants $g_M$ determined by the
$compositeness \, \, \, condition$: that the renormalisation
constants of the meson fields are equal to zero so no {\it elementary meson}:

\begin{equation}
Z_M=1-g^2_M\Pi_M^\prime(m^2_M)=0,
\end{equation}
where the empirical meson mass is the input.
Here $\Pi_M^\prime$ is the derivative of the meson mass operator.  This way
of determining the quark-meson coupling strength is one of the
differences as compared with other approaches using quark loops like
ref.\cite{Ng} in which actual processes like $\pi^0 \to 2\gamma$  is fit to
determine the coupling.
Note that the above equation provides the right normalization
of the charge form factor $F_M(0)=1$. This could be readily seen from
the Ward identity

$$
g^2_M\Pi_M^\prime(p^2)=g^2_M{1\over 2p^2}p^\mu
{\partial \Pi_M(p^2)\over\partial p^\mu}=g^2_M{1\over 2p^2}p^\mu
T^\mu_M(p,p)=F_M(0)=1.
$$
where $T_M^\mu(p,p^\prime)$ is the three-point function  describing the
meson electromagnetic charge
form factor. It may be of use to add that the compositeness condition ensures
that unlike some of the {\it quark-meson hybrid} models, we have no danger of
double counting.

Mesonic interactions (meson-photon, meson-meson, etc.) in QCM proceed only
through  closed quark loops:

\begin{equation}
\int\!\! d\sigma_v {\rm tr}\biggl[M(x_1)S_v(x_1-x_2)...M(x_n)S_v(x_n-x_1)
\biggr].
\end{equation}
Here,

\begin{equation}
S_v(x_1-x_2)=\int\!\!{d^4p\over (2\pi)^4i} e^{-ip(x_1-x_2)}
{1\over v\Lambda-\not\! p}
\end{equation}
is the quark propagator with the scale parameter $\Lambda$ characterising
the size of the confinement (or equivalent to something like the constituent
quark mass), and the measure $d\sigma_v$, which is essential
for quark confinement, is defined to provide the absence of the propagator
singularities in quark loops, prohibiting the physical quark
production:

\begin{equation}
\int\!\!{d\sigma_v\over v-z}\equiv G(z)=a(-z^2)+zb(-z^2).
\end{equation}

The shapes of the confinement functions $a(u)$ and $b(u)$, and the scale
parameter $\Lambda$ have been determined from a reasonable model description
of the low-energy hadronic quantities, see \cite{EI} for more details:

\begin{equation}
a(u)=a_0\exp(-u^2-a_1u) \hspace{1cm} b(u)=b_0\exp(-u^2+b_1u).
\end{equation}
Here we have fixed $a_0 = b_0 =2, a_1 = 1, b_1 = 0.4$ and
$\Lambda=460 \  {\rm MeV}$,
which describe various basic meson constants quite well, some examples of
which may be found in Table 1, for which various intergals defined in Table 2
are used. See also other quantities in \cite{EI}. Note
that in Table 2 and in the rest of this work the dimensionless meson masses
defined as
$\mu_M = m_M /\Lambda$ \ appear. We also note that some of the functions in the
latter table define the meson-quark-antiquark coupling strength through the
compositeness condition in Eq.(10): for example

$$
\frac{3g_\pi^2}{4\pi^2} = \frac{2}{R_{PP}(\mu_\pi^2)}, \hspace{1cm}
\frac{3g_\rho^2}{4\pi^2}=\frac{3}{R_{VV}(\mu_\rho^2)}.
$$

Naturally, the coupling constants $g_{\rho\pi\pi}$,  $g_{\rho\pi\gamma}$, and
$g_{\rho\gamma}$ obtained in the model describe the decay widths of the
$\rho$ successfully, which read
\begin{eqnarray}
\Gamma(\rho &\to& \pi\pi)=m_\rho \frac{g_{\rho\pi\pi}^2}{48\pi}
\biggl[1-\frac{4m_\pi^2}{m_\rho^2}\biggr]^{3/2},\nonumber\\
&&\nonumber\\
\Gamma(\rho &\to& \pi\gamma)=m_\rho \frac{g_{\rho\pi\gamma}^2}{96\pi}
\biggl[1-\frac{m_\pi^2}{m_\rho^2}\biggr]^3, \nonumber\\
&&\nonumber\\
\Gamma(\rho &\to& e^+e^-)=\frac{4\pi\alpha^2}{3}m_\rho g^2_{\rho\gamma},
\nonumber
\end{eqnarray}
Also it may be useful here to mention that from Table 1 and Eq.(5) our model
gives $\alpha_k = 0.53$
(by adopting the standard value: $m_\rho = 770$ MeV), identical to the
empirical value and quite close to the
KSRF value of 1/2.

For our present objective we first calculate the $\pi^0\to\gamma\gamma$
amplitude: $F_{\pi \gamma \gamma}$.  The contribution comes only from
the triangular quark loop (Fig.4): no vector meson pole. From Table 1 one
can read off the expression

\begin{equation}
F_{\pi\gamma\gamma}=\frac{\alpha}{\pi F_\pi}
\biggl[{2R_{PVV}(\mu_\pi^2)R_{P}(\mu_\pi^2)\over R_{PP}(\mu_\pi^2)}\biggr]
=0.94\frac{\alpha}{\pi F_\pi}=0.024 \; {\rm GeV^{-1}},
\end{equation}
in excellent agreement with the experimental value quoted in the introduction
and with the chiral anomaly prediction: $\alpha /\pi F_\pi =0.025 \;
{\rm GeV^{-1}}$ (here again $\mu_M = m_M /\Lambda$ is used).  We note that
in the chiral limit: $m_\pi = 0$, the value in the bracket of Eq.(15)
 becomes 0.97 even somewhat closer to the chiral anomaly
prediction. Thus we may conclude that the model is well calibrated
to have successfully passed the first hurdle.

The next step is to obtain $F_{3\pi}(s,t,u)$  for $\gamma \to 3\pi$
which, in the context of the present model, is described by two types of
diagrams in Fig.5. We will consider an alternative physical
process $\pi\gamma\to\pi\pi$. The invariant matrix element consists of two
parts and reads

\begin{equation}
M^\mu[\pi(p_1)+\gamma(q)\to\pi(p_2)+\pi(p_3)]=
\varepsilon^{\mu \nu \eta \kappa}(p_1)_\nu (p_2)_\eta (p_3)_\kappa
F_{3\pi}(s,t,u),
\end{equation}
where
\begin{eqnarray}
F_{3\pi}(s,t,u)&=&F_{3\pi}^{\rm box}(s,t,u)+ F^{\rm pole}(s,t,u)\\
&&\nonumber\\
F_{3\pi}^{\rm box}(s,t,u)&=&
\frac{e}{\Lambda^3}(\frac{g_\pi}{\sqrt{2}})^3\frac{1}{4\pi^2}
\{R_{\Box}(s,t)+R_{\Box}(t,u)+R_{\Box}(u,s)\}, \\
&&\nonumber\\
F^{\rm pole}(s,t,u)&=& \frac{1}{m_\rho}
\biggl[
\frac{2g_{\rho\pi\pi}(s)g_{\rho\pi\gamma}(s)}{m_\rho^2 -s} +
(s\to t)+(s\to u)
\biggr],
\end{eqnarray}

The function defining the box-diagram (corresponding to the first
diagram in Fig.5 which, however, is not drawn as a {\it proper} box) is
written as

\begin{equation}
R_{\Box}(s,t)=2\int d^4\alpha\delta(1-\sum\limits_{i=1}^4\alpha_i)
\biggl[-a^\prime\biggl(-D_4(\alpha)\biggr)\biggr],
\end{equation}
where $a^\prime$ is the derivative of function $a(u)$ in Eq.(14), and $D_4$ is
equal to

$$
D_4(\alpha)=(\alpha_1\alpha_2+\alpha_2\alpha_3+\alpha_3\alpha_4)
m_\pi^2/\Lambda^2+\alpha_2\alpha_4 s/\Lambda^2+\alpha_1\alpha_3 t/\Lambda^2.
$$

The value of $R_\Box$ changes very slowly: less than 1$\%$  when
s and t are varied within and near the physical region for the process under
consideration.
We may therefore safely approximate the box contribution by setting
$s=t=u=0$ neglecting all the mass and momentum dependence with quite a
good accuracy. Then one trivially finds $R_{\Box}(0,0) = 2$ leading
to
\begin{equation}
F_{3\pi}^{\rm box}(0,0,0)\approx\frac{6R_P(0)}{R_{PP}(0)^3}
\frac{e}{4\pi^2F_\pi^3}=0.22F_{3\pi}^{\rm anom}.
\end{equation}
%

The vector meson pole (or resonance) contributions: the second diagram in
Fig.5, are defined
by $F^{\rm pole}(s,t,u)$ in Eq.(19) which contains the $\rho\pi\pi$ and
$\rho\pi\gamma$ vertices which are {\it not} constants but are explicitly
dependent upon the Mandelstam variables due to their quark loop structure,

\begin{eqnarray}
g_{\rho\pi\pi}(s)&=&2\pi \frac{R_{VPP}(s/\Lambda^2)}
                              {R_{PP}(\mu^2_\pi)}
\sqrt{\frac{2}{R_{VV}(\mu^2_\rho)}},
\nonumber\\
&&\\
g_{\rho\pi\gamma}(s)&=&\frac{e m_\rho}{\Lambda}\sqrt{2\over 3}
\frac{R_{PVV}(s/\Lambda^2)}
     {\sqrt{R_{PP}(\mu^2_\pi)R_{VV}(\mu^2_\rho)}}.
\nonumber
\end{eqnarray}
Fig.6 demonstrates their dependence on the virtual $\rho$ meson mass squared.
Apparently the variations are not quite small. This may provide a rather
interesting effect on which we shall discuss later.
Note that the $\rho$-meson decay widths as presented before are defined by the
 coupling strengths of
Eq.(22) with the $s$-variable set equal to the square of the $\rho$-meson mass:
$g_{\rho\pi\pi}=g_{\rho\pi\pi}(m^2_\rho)$,
$g_{\rho\pi\gamma}=g_{\rho\pi\gamma}(m^2_\rho)$, and
$g_{\rho\gamma}=g_{\rho\gamma}(m^2_\rho)$.
As is well known, there are relations between these constants (see, for
example,
\cite{Saku})

\begin{equation}
g_{\rho\gamma}=1/g_{\rho\pi\pi}, \hspace{1cm}
g_{\rho\pi\gamma}=\frac{1}{2}g_{\rho\pi\pi}\cdot
\frac{m_\rho}{e}F_{\pi\gamma\gamma}=
em_\rho\frac{g_{\rho\pi\pi}}{8\pi^2F_\pi},
\end{equation}
which are obtained based upon an assumption on the VMD
and PCAC relation.
In our model the corresponding relations are found numerically as

\begin{equation}
g_{\rho\gamma}=1.08/g_{\rho\pi\pi}, \hspace{1cm}
g_{\rho\pi\gamma}=\frac{0.89}{2}g_{\rho\pi\pi}\cdot
\frac{m_\rho}{e}F_{\pi\gamma\gamma}=
0.84em_\rho\frac{g_{\rho\pi\pi}}{8\pi^2F_\pi},
\end{equation}
where our result in Eq.(15) has been taken into account. One can see that the
relations (23) are reproduced
within the commonly accepted accuracy of the VMD assumption although our
model is not based upon it. In this respect
we should stress that with empirical values for $g_{\rho \pi \pi}$ etc. one
obtains the result very similar to ours in Eq.(24).

Upon combining the box and $\rho$-pole contributions, we find for null momenta
(s=t=u=0, soft pions and photon)

\begin{eqnarray}
F_{3\pi}(0)& \equiv & F_{3\pi}(0,0,0)=
F_{3\pi}^{\rm box}(0,0,0)+
\frac{6g_{\rho\pi\pi}(0)g_{\rho\pi\gamma}(0)}{m^3_\rho}\\
&&\nonumber\\
&=&\frac{e}{4\pi^2F_\pi^3}[0.22+0.83]=1.05F_{3\pi}^{\rm anom}, \nonumber
\end{eqnarray}
which is very close to the chiral anomaly prediction
(the ATA derivation) or, equivalently, to the one from the
electromagnetically gauged Wess-Zumino-Witten term as discussed in the
previous section although we have borrowed nothing from those approaches.

A particularly noticeable feature of our result is the dominant r\^ ole
played by the $\rho$ meson pole contribution, similar to the VMD
approaches. In this respect, it may be useful to make here a close comparison
of our result with the VMD result.  Within the VMD models (see, for example)
\cite{Rudaz} the zero
energy $\gamma \to 3\pi$ amplitude from the vector meson pole contribution
may be obtained as
\begin{equation}
F^{\rm VM}_{3\pi}(0) = 3\alpha_k\cdot F^{\rm anom}_{3\pi},
\end{equation}
the derivation of which requires the relations in Eq.(23).  A possible
presence of the direct contact term \cite{Rud2} introduces an additional
contribution of the form

\begin{equation}
F^{\rm con}_{3\pi}(0) = \gamma(\alpha_k)\cdot F^{\rm anom}_{3\pi},
\end{equation}
thus
\begin{equation}
F_{3\pi}(0) = F^{\rm con}_{3\pi} + F^{\rm VM}_{3\pi} =
[\gamma(\alpha_k)+ 3\alpha_k]\cdot F^{\rm anom}_{3\pi}.
\end{equation}
While the form of the $\rho$ meson pole contribution is identical in any
VMD approaches, that of the contact contribution differs in different models:
for example, Refs. \cite{Pak} \cite{Rud2} find $\gamma(\alpha_k) =
1-3\alpha_k$,
whereas \cite{Kay} obtained $\gamma(\alpha_k) = 1-3\alpha_k + 3/2\alpha_k^2$
(we
recall that the first form works well with the ATA theorem while the second
can be made consistent with the $\omega \to 3\pi$ width, both with
{\it reasonable} values of $\alpha_k$).

Regarding our result Eq.(25), it may appear quite natural to identify the box
contribution as that coming from the contact term in the effective meson
Lagrangian. Then although, as we repeat, our model is not based in any way upon
the VMD assumption, it would be tempting to assume (although somewhat dubious
!)
that the relations in
Eq.(23) may well hold among the coupling strengths evaluated at zero energy,
viz. among $g_{\rho \pi \pi}(0)$, $g_{\rho \pi \gamma}(0)$, etc. Then from
Eqs.(25-28) this seemingly natural but somewhat naive assumption leads to
$\alpha_k = 0.28$ (from $3\alpha_k =0.83$). We note that since the form of the
contact term is not unique, it is wiser not to use it to extract $\alpha_k$.
Clearly, the thus obtained value of $\alpha_k$ appears quite smaller as
compared with both the empirical and the KSRF values. Does this imply that our
model is not consistent with the Vector Dominance Model quantitatively, or
qualitatively, or even is there anything wrong with our approach?

In order to investigate this problem, we want to stress that the key point is
the fact that, as is clear in Eq.(22) in our quark approach, the couplings to
the  $\rho$ meson are functions of the virtual mass square of the $\rho$, and
that , as stated before, the mass square dependence is not small (see Fig.6).
As a  consequence, the $\rho$-meson contribution in Eq.(25) evaluated at zero
$\rho$ mass square is considerably smaller than the ordinary value obtained in
VMD models with fixed coupling strengths (corresponding to those strength
functions evaluated at the physical $\rho$ mass). This then leads to the
extraction of small $\alpha_k$ mentioned above as long as we do not
queation the appropriateness of such an extraction and  identification as
$\alpha_k$.  Now if one insists that somehow
a consistency with VMD is mandatory in order our approach to be acceptable, it
may be possible to reinterpret our result for that objective as follows:
Eq.(19) which is the $\rho$-pole
contribution to the $\gamma \to 3\pi$ amplitude in our model may be
decomposed into the {\it pure} $\rho$ meson pole and the rest, the latter being
analytic at the $\rho$ meson pole,
\begin{equation}
\frac{2\cdot g_{\rho \pi \pi}(m_\rho^2)g_{\rho \pi \gamma}(m_\rho^2)}{m_\rho (
m_{\rho^2} -s)} + R(s) + (s \to t) + (s \to u).
\end{equation}
The {\it pure} $\rho$-pole contribution in the above expression
is the one appearing as the $\rho$ meson pole contribution in the
VMD models.  The analytic part is beyond the VMD approach. One may then add the
analytic part to the box contribution and regard  the sum as the contact term
introduced phenomenologically in the effective chiral vector meson
Lagrangian approaches stated in the last section. With this
rearrangement, Eq.(25) may be rewritten as

\begin{equation}
F_{3\pi}(0)=F_{3\pi}^{\rm anom}
\biggl(
F^{\rm con}+
3\cdot\frac{2g_{\rho\pi\pi}(m^2_\rho)g_{\rho\pi\gamma}(m^2_\rho)}
{m^3_\rho F_{3\pi}^{\rm anom}}
\biggr),
\end{equation}
where the first term in the above bracket is the contact term contribution
defined as
\begin{equation}
F^{\rm con}=\frac{F_{3\pi}^{\rm box}(0,0,0)}{F_{3\pi}^{\rm anom}}-
3\cdot\frac{2[g_{\rho\pi\pi}(m^2_\rho)g_{\rho\pi\gamma}(m^2_\rho)-
         g_{\rho\pi\pi}(0)g_{\rho\pi\gamma}(0)]}
{m^3_\rho F_{3\pi}^{\rm anom}},
\end{equation}
whereas the second term is the standard $\rho$ pole contribution. It is
easy to show that this second term may be identified as $3\alpha_k$ of
Eq.(5) when
relations in Eq.(23) are strictly obeyed. In our model where, instead,
 Eq.(24) holds, the effective value of $\alpha_k$ is obtained as
\begin{equation}
\alpha_k^{eff}=\frac{2g_{\rho\pi\pi}(m^2_\rho)g_{\rho\pi\gamma}(m^2_\rho)}
{m^3_\rho F_{3\pi}^{\rm anom}}=
0.84\cdot \biggl(
                 \frac {g_{\rho \pi \pi} F_\pi} {m_\rho}
           \biggr)^2_{\rm empirical}=0.44.
\end{equation}
Clearly, this is consistent with the KSFR prediction
($\alpha_k=0.5$) within the accuracy of the VMD approach.
Then from Eq.(31) the corresponding {\it contact term} is equal to
$F^{\rm con}=1.05-3\alpha_k^{eff}=-0.27$, which is consistent with the one
in effective Lagrangian approaches both in sign and magnitude. Thus our result
has been shown not in conflict with the VMD models.

Our amplitude with non-vanishing Mandelstam variables  may be conveniently
re-written in terms of $F_{3\pi}^{\rm anom}$,

\begin{equation}
F_{3\pi}(s,t,u)=F_{3\pi}^{\rm anom}
\biggr[0.22+C_\rho(s)D_\rho(s)+C_\rho(t)D_\rho(t)+C_\rho(u)D_\rho (u)\biggr],
\end{equation}
where
$$
C_\rho(s)=\frac{2g_{\rho \pi \pi}(s)g_{\rho \pi \gamma}(s)}{m_\rho^3F_
{3\pi}^{\rm anom}}, \hspace{0.5cm} {\rm and} \hspace{0.5cm}
D_\rho (s)=\frac{m_\rho^2}{m_\rho^2-s}.
$$

\section{Comparison of Models With Data from Existing Primakoff Process}

Here we first discuss the value of the zero energy $\gamma \to 3\pi$
amplitude from various models.
As already discussed, the chiral anomaly prediction is $9.5$  (in units of
${\rm GeV^{-3}}$).
Of those models incorporating vector mesons, some have reproduced (or have
been taylored to reproduce) the chiral anomaly value \cite{Pak} \cite{Rud2}
\cite{Kura} while some other \cite{Kay} gives a larger value: about $13.0$
due to the smaller magnitude of the $\omega 3\pi$ contact term. The chiral
perturbation calculation of ref.\cite{Bij}  to $O(p^6)$ as quoted by
Moinester \cite{Moin} offers the number $10.7 \pm 0.5$ (it should be mentioned
here that at the zero energy [or the soft pion limit] this approach should
reproduce the anomaly prediction so the quoted value should be taken with
some caution). Our QCM yields
$10.0$ for its value. These values should be compared with the one
extracted  from the Primakoff process \cite{Antip} is  $12.9 \pm 0.9 \;
{\rm (stat)} \pm 0.5 \; {\rm (sys)}$. Based upon this experimental value
it is hard to rule out (or give preference to) any of the prediction.

Next, the model amplitude will be used to predict the integrated Primakoff
cross section for the pion production by incoming pions as in Fig.1.
We adopt the Mandelstam variables as introduced previously and go to the
centre of mass system for the two final state pions. Then
for $\gamma\pi\to \pi\pi$ (see, Fig.1) noting that $s+t+u=3m_\pi^2 + q^2$,
we have
$$
\vec p_1+\vec q=\vec p_2+\vec p_3=0, \hspace{1cm}
\cos\theta=\frac{(\vec p_1\vec p_2)}{|\vec p_1||\vec p_2|}, \hspace{1cm}
p_1^0=E,
$$

and taking into account that the photon is almost real
$(q^2\approx 0)$, one may write

$$
t=\frac{1}{2}[3m_\pi^2-s+(s-m_\pi^2)\sqrt{1-4m_\pi^2/s}\cos\theta],
$$

$$
u=\frac{1}{2}[3m_\pi^2-s-(s-m_\pi^2)\sqrt{1-4m_\pi^2/s}\cos\theta].
$$

The physical region for $s$ is $4m_\pi^2\le s \le 16 m_\pi^2$.
We obtain various model amplitudes for
$F_{3\pi}(s,t,u)$ in units of the chiral anomaly prediction:
$F_{3\pi}^{\rm anom} \equiv e/4\pi^2F_\pi^3$ as a function of $s$ for
the forward scattering angle: Fig.7
(the angular dependence has been found rather weak so we only give only
one angular value).   The plotted curves are the original Rudaz VMD model
result
\cite{Rudaz}, then his modified (improved)
 amplitude
\cite{Rud2}, the amplitude from Kaymakcalan {\it et al.}'s model \cite{Kay},
chiral perturbation (CHPT) prediction \cite{Bij}, and our present quark
model result.
Except for the amplitudes from \cite{Kay} and \cite{Bij}, other
curves are mutually rather close. Kaymakcalan {\it et al.}'s \cite{Kay}
result
is due to its smaller sized of the contact term which has given a large zero
energy amplitude.
On the other hand the CHPT amplitude above
$s \ge 8m_\pi^2$ has come out quite differently from the others.
The validity for the CHPT expansion is up to about this value of $s$,
so the resultant difference may be natural. If future experiments may be
able to extract this $s$ and scattering angle
dependent behaviour of the amplitude, it would be of quite help. At present,
however, we should be satisfied with their prediction on the available
data \cite{Antip}.

The cross section for pion production by the pion incident in the nuclear
Coulomb field (the Primakoff process)
\begin{equation}
\pi^-+(Z,A)\to \pi^-+\pi^0+(Z,A)
\end{equation}
as in Fig.1 may be expressed in terms of the $\gamma \pi \to 2\pi$
cross section through the equivalent photon method:

\begin{equation}
\frac{d\sigma}{dsdtdq^2}=\frac{Z^2\alpha}{\pi}
\biggl[\frac{q^2-q^2_{\rm min}}{q^4}\biggr]
\frac{1}{s-m_\pi^2}\frac{d\sigma_{\gamma\pi\to\pi\pi}}{dt},
\end{equation}
where

$$
q^2_{\rm min}=\biggl(\frac{s-m^2_\pi}{2E}\biggr)^2,
$$
and

\begin{equation}
\frac{d\sigma_{\gamma\pi\to\pi\pi}}{dt}=
\frac{|F_{3\pi}(s,t,u)|^2}{512\pi}
(s-4m_\pi^2)\sin^2\theta
\end{equation}
with the vertex $F_{3\pi}(s,t,u)$ defined by the dynamics of the
$\gamma\to 3\pi$ transtion.

Neglecting the $q^2$-dependence in $F_{3\pi}(s,t,u)$, viz.
setting $q^2 \approx 0$ (thus $s+t+u=3m_\pi^2$), the integrated cross section
per unit proton charge in the nucleus is given as

\begin{equation}
\frac{\sigma}{Z^2}=\frac{\alpha}{1024\pi^2}
\int\limits_{4m_\pi^2}^{s_{\rm max}}ds
\frac{(s-4m_\pi^2)^{3/2}}{\sqrt{s}}
\biggl\{\ln{\frac{q^2_{\rm max}}{q^2_{\rm min}}}-1+
\frac{q^2_{\rm min}}{q^2_{\rm max}}\biggr\}
\int\limits_0^\pi d\theta \sin^3\theta |F_{3\pi}(s,t,u)|^2.
\end{equation}

The condition for the Primakoff effect is the following \cite{Teren}
\cite{Antip}

$$
E^2>>m_\pi^2\sim s>>q^2.
$$

This condition is better satisfied when the energy of the incident pion
beam becomes higher. The Serpukhov experiment \cite{Antip} has been carried out
with the 40 GeV pion beam. The maximum momentum transfer was
$q^2_{\rm max}=2\cdot 10^{-3} \, {\rm GeV}^2$
and $s_{\rm max}=10m_\pi^2$.

The result of the calculation of the cross section for various approaches are
shown
in Table 3. One can see that the results are rather close to each other,
except for the one from the model of ref. \cite{Kay}, which
could be understood because its prediction for $F_{3\pi}(0)$ is about
30$\%$
larger than the chiral anomaly prediction, as discussed in Section 3.
We note that a somewhat larger value (about 10$\%$) of the cross
section in our QCM
model as compared with those from those of refs. \cite{Rud2} and \cite{Bij}
is also largely the reflection that our $F_{3\pi}(0)$ is about 5$\%$
larger
than the chiral anomaly prediction.  In Table 3  the row
denoted as {\it Chiral Anomaly} has been obtained from
replacing $F_{3\pi}(s,t,u)$ simply by $F_{3\pi}^{\rm anom} \equiv
e/4\pi^2 F_\pi^3$, thus disregarding the energy dependence (and
 angular variations) in calculating
$\sigma/Z^2$ above. This is found to reduce the integrated cross section
by more than 10$\%$. But at present, the extracted values from the
experiment may only be able to say that the
result from ref.\cite{Kay} may be inadequate (note, however, that as we
have stated before, one K-meson loop correction seems to have restored the
value of the contact term to the Rudaz one \cite{Rud2}).  Improved experiments
are definitely in need.

\section{Discussion and Conclusion}
\baselineskip 20pt
It may sound repetitive, but we make a pedagogical review at the
beginning of the discussion.

Only the {\it anomalous} (but not the ordinary) PCAC combined with current
algebra was able
to correctly predict the observed $\pi^0 \to 2\gamma$ decay rate
\cite{Adletet}, and
the zero energy $\gamma \to 3\pi$ amplitude $F_{3\pi}(0)$ was predicted from
the former
amplitude: we have termed this as the ATA theorem \cite{Adler} \cite{Teren}
\cite{Aviv}. This theorem was rederived quite naturally later from the
electromagnetically gauged Wess-Zumino-Witten action (WZW) \cite{Wess}
\cite{Witten}. Modern versions of the vector dominance models (VMD)
 \cite{Kay} \cite{Pak} \cite{Rud2} \cite{Kura} were
used in an attempt to reach this theorem with the help of the KSRF relation
\cite{KSRF} (and its more relaxed form). In order to be consistent with the
ATA theorem the vector dominance models have been required to introduce
a non-zero constant four-point coupling for $\omega \to 3\pi$ leading
to a contact $\gamma \to 3\pi$ coupling which has an opposite sign as
compared with the vector meson pole contribution. While the vector meson
pole contribution in these models for the zero energy $\gamma \to 3\pi$
amplitude is $1.5$ times the value obtained from the ATA theorem, the value
of the contact term contribution differs: - 1/2 times the ATA value from refs.
\cite{Pak} \cite{Rud2} \cite{Kura}, and about 4 times smaller in ref.\cite
{Kay} (when evaluated at the tree level), and the former is only consistent
with the ATA theorem.
We have calculated above the Primakoff cross section
for these models and found that improved experiments may be able to
say which group of the VMD model (regarding the contact term value) should be
more relevant to the $\gamma \to 3\pi$ amplitude.

We have devised a confined quark model (QCM) to study the same subject
without implementing any explicit chiral anomaly. The result has come out
quite consistent with the ATA theorem and the
consequence of the electromagnetically gauged WZW action. The obtained
amplitude
$F_{3\pi}(s,t,u)$ is fairly close to the VMD prediction of
refs.\cite{Pak} \cite{Rud2}
 \cite{Kura}. However, if we regard our quark
box diagram as the contact term contribution, it is found to have the same
sign as the vector meson pole contribution, and
counts just 22$\%$ of the ATA
theorem (or the chiral anomaly) prediction:
$F_{3\pi}^{\rm anom} \equiv e/4\pi^2F_\pi^3$. This then implies (as it has
eventually turned out)
that our vector meson pole contribution accounts for about 80$\%$ of
the chiral anomaly prediction. This should be contrasted to the result of the
VMD prediction discussed above which provides 1.5$F_{3\pi}^{\rm anom}$
upon emplying the KSRF relation: $\alpha_k = 1/2$. At this point we want to
remark
 that, as in Fig.6 both our $\rho \pi \pi$ and $\rho \gamma \pi$ coupling
vertices are functions of the $\rho$ meson mass squared arising from the
quark loop, and that their variations are not negligible, especially the
former. Their values at $m_\rho^2$ are practically the same as those in the
VMD models, see Table 1, but at zero mass they are smaller. Those smaller
values have been used to find the value of $\approx 80\%$ for the
$\rho$-pole contribution mentioned above.
When those
coupling constants are evaluated
at $m_\rho^2$, one finds that the vector meson contribution to $F_{3\pi}(0)$
becomes 1.3 times the chiral anomaly prediction, not very different from the
VMD result of 1.5 times (recall Eqs.(30,32)).

As for the difference in sign and magnitude of the contact term between our
quark model prediction and those from the VMD models, it does not appear easy
to
find the answer to this discrepancy {\it as long as we stick to regarding the
the contact term as our box contribution}: within our confined quark model
approach the loop
integral for the box diagram is completely convergent and gauge invariant.
In the approach in ref.\cite{Aviv}, the loop integral generates a constant
term breaking the gauge invariance so it must be subtracted out (it is related
to the so-called surface term arising from the shifting of the integration
variable
in a divergent integral, see for example section 6.2 of ref.\cite{Cheng}). This
subtraction appears to be the origin of the sign change which we do
not have in QCM.  So what could be the possible compromise ?
There may be some effects coming from the excited vector meson(s) like
$\rho'$ which was expected to cure the trouble with the $\omega \to 3\pi$
rate with the larger contact term: $-0.5F_{3\pi}^{\rm anom}$
\cite{Kramer}. Or the effect of the
Regge trajectory contributions might be included \cite{Kura}. Either of these
effects might be included in the vector meson pole contributions. These lines
of
considerations might work as a bridge between our and the VMD
pictures: one might suspect that some type of hadron-quark duality (for
example,
quark loops vs. meson loops) be operative somewhere.

Once we adopt a somewhat more relaxed definition of our contact term, however,
the resolution of the apparent discrepancy in sign and the size of the contact
term in our and VMD models may seem feasible. This could proceed along the line
we exploited in Section 3 when extracting $\alpha_k^{eff}$ from our prediction
of $F_{3\pi}(0)$. Namely, we decompose our $\rho$-pole contribution into
the {\it pure} pole and analytic parts. Then the analytic part is added to the
box contribution which one may regard as the effective contact term. Now, this
effective contact term is no longer a mere constant, but a function of the
Mandelstam variables: $s$, $t$ and $u$. It is then easy to see, as found
in Section 3, that in the zero energy limit this term obtains a negative sign
with magnitude about half the Rudaz value \cite{Rudaz} making our $F_{3\pi}(0)$
prediction very close to the chiral anomaly one . At finite energies
(finite $s$)
one may be able to see that the magnitude of this term decreases, and this
is consistent with the apparent need for small (or zero) contect term
contribution in the $\omega \to 3\pi$ width within the context of the
effective chiral vector meson models. We note in this respect that both
experimental and model investigations, for
example, on
$e^+e^- \to 3\pi$ (including $\omega \to 3\pi$) should be quite helpful.

{}From the point of view of the
proposed $\gamma \pi \to \pi \pi$ experiments (either the photon or pion
in the initial state is virtual, or at best close to $real$) \cite{Moin}
\cite{CEBAF} to extract $F_{3\pi}(0)$ or more ambitiously to deduce
$F_{3\pi}(s,t,u)$ for different $s$ and c.m. scattering angle (for example,
to clarify the
distinction between our quark model and VMD predictions: notably the r\^ ole
played by either the constant or non-constant contact term, the latter
arising from the Mandelstam variable dependent $\rho$ couplings), they will be
required to attain a very high precision in data acquisition and analyses.
In this respect it may be useful to be reminded that if some sort of analytic
extrapolation to a pole (either for the pion or photon) is needed in analysing
the data, one must not forget
the fact that assumed analyticity
property (of the relevant amplitude) and the statistical nature of the data
are mutually exclusive. So instead of a na\" {\i}ve straightforward
extrapolation,
some optimal mapping may have to be sought beforehand in order to minimise
the unwanted (and in most cases the dominant) background contributions,
see for example refs.\cite{Ciulli} \cite{Locher} concerning this procedure.
Finally, in
view of the closeness of the predictions from some VMD models, based upon
the chiral plus the WZW action which basically implements the chiral anomaly,
and our confined quark model approach (which, at least, is not explicitly
built upon chiral anomaly like the WZW term), may it be of some use to ask
ourselves if the proposed experiments will be really testing
the ATA theorem as a consequence of chiral anomaly ?

\begin{center}
{\bf ACKNOWLEDGMENTS}
\end{center}

We would like to acknowledge helpful discussions with E.A. Kuraev
on the question of the contact term in effective meson Lagrangian approaches.
T.M. is greatful  to Professor D.V. Shirkov and the members of the
Laboratory of Theoretical Physics, JINR, Dubna for their warm hospitality.
The manuscript was completed while T.M. was enjoying a kind hospitality  at the
Inst. de Physique Nucl\'eaire, Lyon, France.  He is grateful to Jean Delorme
for critical reading of the manuscript.
This work was supported in part by the United States Department of Energy
under Grant No. DE-FG-ER40413. A support for M.A.I. by Soros Foundation
under Grant No. SDF000 is greatly appreciated.

\newpage
\listoffigures

\noindent
Fig.1. Pion production by an incident pion in the Coulomb field of a
heavy nucleus (the Primakoff process) \cite{Antip}, \cite{Moin}.

\bigskip
\noindent
Fig.2. Pion electroproduction by an incident pion \cite{Amend}.

\bigskip
\noindent
Fig.3. Photo double pion production on the proton \cite{CEBAF}.

\bigskip
\noindent
Fig.4. The $\pi^0 \to 2\gamma$ decay through the quark triangle.

\bigskip
\noindent
Fig.5. The $\gamma \to 3\pi$ in QCM. The first diagram is called the
$Box$ contribution while the second type is termed as the vector meson pole
contribution.

\bigskip
\noindent
Fig.6. The squared mass dependence of the $\rho \to \pi\gamma$ and
$\rho \to \pi\pi$ coupling vertices in QCM normalised to the values at the
$\rho$ mass.

\bigskip
\noindent
Fig.7. The $\gamma \to 3\pi$ amplitude for the physical process
$\gamma \pi \to 2\gamma$  for zero scattering angle, normalised to the chiral
anomaly prediction: $F_{3\pi}^{\rm anom} \equiv  e/4\pi F_\pi^3$.
Various model predictions drawn are (1) the VMD model of Rudaz \cite{Rudaz},
(2) the modified Rudaz model implementing the KSRF relation \cite{Rud2},
(3) chiral Lagrangian with Wess-Zumino-Witten term gauged by massive
vector-axial vector Yang-Mills fields \cite{Kay}, (4) the Chiral
Perturbation calculation of Bijnens $et\  al.$ ref.\cite{Bij} (since the
amplitude is complex, its absolute value is plotted for this approach),
and (5) the
present confined quark model (QCM).
\newpage
\listoftables

\noindent
Table 1. Some QCM results for low energy mesonic quantities as
compared with their experimental counterparts.

\bigskip
\noindent
Table 2. Various functions entering the QCM calculation of meson
verticies (coupling strengths), masses (self energies), etc. Functions
$a(u)$ and $b(u)$ characterise the confined quark propagator as discussed in
Section 3.

\bigskip
\noindent
Table 3. Comparison of several model predictions and the experimental
result of the cross section for $\pi^- \gamma \to \pi^- \pi^0$.  Experiment
was done in the Coulomb field of a nucleus of charge $Z$ (Primakoff effect),
thus the appropriate quantity to be compared with model results is
$\sigma/Z^2$.
\newpage

\begin{center}
Table 1. Several low energies quantities obtained in QCM.
\end{center}

\large
\begin{center}
\def\arraystretch{2.0}
\begin{tabular}{|l|l|l|}
\hline\hline
         & {\rm QCM} & {\rm Expt.} \\
\hline\hline
$F_\pi={\Lambda R_P(\mu^2_\pi)\over 2\pi}\sqrt{3\over R_{PP}(\mu^2_\pi)}$ &
93 {\rm MeV} & 93 {\rm MeV} \\
\hline
$F_K={\Lambda R_P(\mu^2_K)\over 2\pi}\sqrt{3\over R_{PP}(\mu^2_K)}$ &
112 {\rm MeV} & 111 {\rm MeV} \\
\hline
$F_{\pi\gamma\gamma}={e^2\over\pi}{1\over\Lambda}
{R_{PVV}(\mu^2_\pi)\over\sqrt{3 R_{PP}(\mu^2_\pi)}}$ &
0.024 ${\rm GeV^{-1}}$ & 0.025 ${\rm GeV^{-1}}$ \\
\hline
$g_{\rho\pi\pi}={2\pi\sqrt{2}R_{VPP}(\mu^2_\rho)\over
R_{PP}(\mu^2_\pi)\sqrt{R_{VV}(\mu^2_\rho)}}$ &
6.0 & 6.0\\
\hline
$g_{\rho\pi\gamma}={e m_\rho\over\Lambda}\sqrt{2\over 3}
{R_{PVV}(\mu^2_\rho)\over\sqrt{R_{PP}(\mu^2_\pi)R_{VV}(\mu^2_\rho)}}$ &
0.16 & 0.17 \\
\hline
$g_{\rho\gamma}={1\over 2\pi}
{R_{V}(\mu^2_\rho)\over\sqrt{2R_{VV}(\mu^2_\rho)}}$ &
0.18 & 0.20 \\
\hline\hline
\end{tabular}
\end{center}

\normalsize
\begin{center}
Table 2. Structural integrals.
\end{center}

\large
\begin{center}
\def\arraystretch{2.0}
\begin{tabular}{|c|}
\hline\hline
$R_{PP}(x)=B_0+\frac{x}{4}\int\limits_0^1dub(-\frac{xu}{4})
\frac{(1-u/2)}{\sqrt{1-u}}$\\
\hline
$R_{VV}(x)=B_0+\frac{x}{4}\int\limits_0^1dub(-\frac{xu}{4})
\frac{(1-u/2+u^2/4)}{\sqrt{1-u}}$\\
\hline
$R_P(x)=A_0+\frac{x}{4}\int\limits_0^1dua(-\frac{xu}{4})\sqrt{1-u}$\\
\hline
$R_{PVV}(x)=\frac{1}{4}\int\limits_0^1du a(-\frac{xu}{4})
\ln\biggl({\frac{1+\sqrt{1-u}}{1-\sqrt{1-u}}}\biggr)$\\
\hline
$R_{VPP}(x)=B_0+\frac{x}{4}\int\limits_0^1dub(-\frac{xu}{4})\sqrt{1-u}$\\
\hline
$R_{V}(x)=B_0+\frac{x}{4}\int\limits_0^1dub(-\frac{xu}{4})
(1+u/2)\sqrt{1-u}$\\
\hline
$A_0=\int\limits_0^\infty dua(u)=1.09$ \hspace{1cm}
$B_0=\int\limits_0^\infty dub(u)=2.26$ \\
\hline\hline
\end{tabular}
\end{center}

\newpage

\large

\begin{center}
{\large Table 3.\\}
\end{center}
\def\arraystretch{2.0}
\begin{center}
\begin{tabular}{|l|l|l|}
\hline\hline
\multicolumn{3}{|c|}{ $\sigma/Z^2$ \, {\rm (nb)} }\\
\hline
\multicolumn{2}{|c|}  {\rm Model}       &    {\rm Expt.} \cite{Antip} \\
\hline\hline
{\rm Chiral anomaly} & 0.95 &             \\
\hline
{\rm Modified Rudaz} \cite{Rud2} & 1.17  & $1.78\pm 0.47 \;$ {\rm (C)}   \\
\hline
{\rm Bijnens} {\em et al.} \cite{Bij} & 1.12  & $1.54\pm 0.34 \;$ {\rm (Al)}
 \\
\hline
{\rm Kaymacalan}  {\em et al.} \cite{Kay} & 2.10   & $1.64\pm 0.37 \;$
{\rm (Fe)}  \\
\hline
{\rm QCM} & 1.27 &\\
\hline\hline
\end{tabular}
\end{center}

\end{document}